\documentclass{elsart3}
\usepackage{graphics}
\usepackage{epsfig}
\usepackage{amssymb}
\usepackage{cite}
\usepackage{subfigure}

\begin{document}

\begin{frontmatter}




\title{{\bf Luminescence quenching of the triplet excimer \\
 state by air traces  in gaseous argon}}

\author[uniz]{C.~Amsler\corauthref{cor1}} \ead{claude.amsler@cern.ch},
\author[uniz]{V.~Boccone},
\author[uniz]{A.~B\"uchler},
\author[eth]{ R. Chandrasekharan},
\author[uniz]{C.~Regenfus}, 
\author[uniz]{J.~Rochet}

\corauth[cor1]{Corresponding author}
\address[uniz]{Physics Institute, University of Z\"urich,  CH--8057 
Z\"urich, Switzerland}
\address[eth]{Institute for Particle Physics, ETH-Z\"urich, CH--8093 Z\"urich, Switzerland}
\begin{abstract}
While developing a liquid argon detector for dark matter searches we  investigate   the influence of air contamination  on the VUV scintillation yield in gaseous argon at atmospheric pressure. We determine with a radioactive $\alpha$-source the photon yield for various partial air pressures and different reflectors and wavelength shifters. We find for the fast scintillation component a time constant $\tau_1$ = 11.3 $\pm$ 2.8 ns, independent of gas purity. However,  the decay time of the slow component depends on gas purity and is a good indicator for the total VUV light yield. This dependence is attributed to impurities destroying  the long-lived  argon excimer states. The population ratio between the slowly and the fast decaying excimer states is determined for $\alpha$-particles to be 5.5 $\pm$ 0.6 in  argon gas at 1100 mbar and room temperature. The measured mean life of the slow component is $\tau_2$ = 3.140 $\pm$ 0.067 $\mu$s at a partial air pressure of 2 $\times$ $10^{-6}$ mbar. 

\end{abstract}


\end{frontmatter}

\section{Introduction}
\label{intro}
Noble liquids such as argon (or xenon) can act as targets for WIMPs (Weak Interacting Massive Particles), the most popular candidates for dark matter in the universe. These elements have high scintillation yields and are also suitable for charge detection because of their relatively low ionization potentials. Both ionization and scintillation light can be detected \cite{ARWarp,ARRubbia}. Argon ($^{40}$Ar) is cheap compared to xenon and is therefore competitive  for large volumes, in spite of its contamination by the  $^{39}$Ar $\beta$-emitter. 
Here we present measurements on gaseous argon done while developing the scintillation light read out of a 1 ton  liquid argon TPC  to search for dark matter (ArDM, \cite{ARRubbia}). 

\begin{figure}[htb]
 \centering
    \includegraphics[width=70mm]{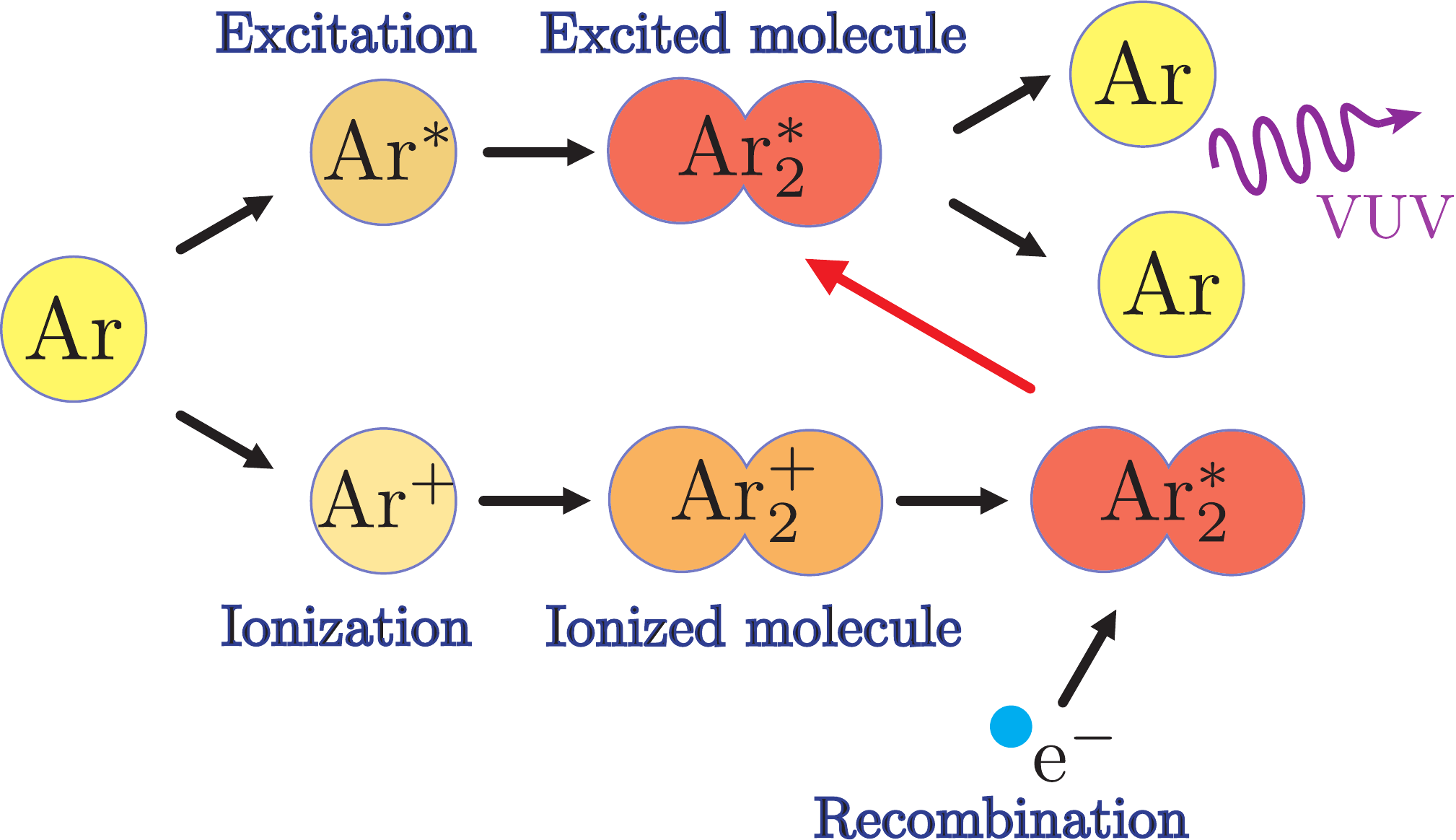}
    \caption{\small\it The two mechanisms leading to the emission of 128 nm photons (adapted from ref. \cite{suzuki}). }
    \label{Aremission}
\end{figure}

The light yield and the mechanism for the luminescence of noble gases and liquids are comparable to that of alkali halide crystals \cite{jortner,cheshnovsky} and are described in the literature for dense gases \cite{keto,suzuki} and liquids \cite{kubonaka,kubohishi,hitachi,doke}. Fundamental to the scintillation process is the formation of excited dimers (excimers) which decay radiatively to the unbound ground state of two argon atoms. Fi\-gure \ref{Aremission} shows schematically the two mechanisms leading to light emission  in argon, excitation and ionization. Excitation leads through collisions with neighbouring atoms to neutral excimers Ar$_2^*$ which decay radiatively into two argon atoms. Ionization leads to the formation of   charged excimers which are neutralized by  thermalized electrons (recombination luminescence). Both processes are strongly pressure and density dependent. For gaseous argon at room temperature and normal pressure, at which we operate here, excitation dominates  \cite{Suzuki2,Carvalho}, while recombination luminescence becomes important at high pressures or  in  liquid.

The argon excimers are created in three nearly degenerate spin states, two  singlets ($^{1}\Sigma_{u}^-$ and $^{1}\Sigma_{u}^+$) and a triplet ($^{3}\Sigma_{u}^+$). The  $^{1}\Sigma_{u}^-$ state cannot decay radiatively by parity conservation, and the  $^{3}\Sigma_{u}^+$ is expected to have a much longer lifetime than the $^{1}\Sigma_{u}^+$ since it has to decay into two spin-0 argon atoms. The  $^{3}\Sigma_{u}^+$ and $^{3}\Sigma_{u}^+$ states decay radiatively by emitting VUV photons  in a $\approx$10 nm band around 128 nm. These photons are not absorbed by atomic argon and can therefore  be detected. Light at higher wavelengths (mostly in the near  IR) is also produced from transitions between highly excited argon atomic states.

The production times of the  triplet and singlet states and their production ratio vary with argon density and also depend on the type of projectile, e.g. electron, $\alpha$-particle or fission fragment \cite{hitachi}. However, the decay times are not affected. 

The time constants of the singlet and triplet states have been measured in dense gases with 160 keV electrons \cite{keto}. The production time of the singlet excimer is around 40 ns at 3 atm, decreasing with increasing gas pressure, and the mean life is about 4 ns.  The mean life of the triplet excimer state is substantially larger, 3.2 $\pm$ 0.3 $\mu$s. 

In liquid argon the mean lives scatter in the range  between 4 and 7 ns for the singlet state, and  between 1.0 and 1.7 $\mu$s for the triplet state. The triplet to singlet production ratios  are  0.3, 1.3 and 3 for electrons, $\alpha$-particles and fission fragments, respectively (for a compilation see ref. \cite{hitachi}). The  large difference in time constants between singlet and triplet states is unique for argon among noble gases and can be used to suppress background in WIMP searches. 

In this paper we present evidence for the (non-radiative) destruction of triplet states in gaseous argon by  traces of air \cite{angie,regi}. The population of the triplet states decreases exponentially with a shorter decay time than measured  with clean argon gas. The reduced  light yield is presumably due to  collisional destruction of the long-lived triplet state by impurities such as water molecules. A similar effect is observed in liquid argon \cite{ArDMliq}.

\section{Experimental setup}
\label{setup}

\begin{figure}[htb]
 \centering
    \includegraphics[width=50mm]{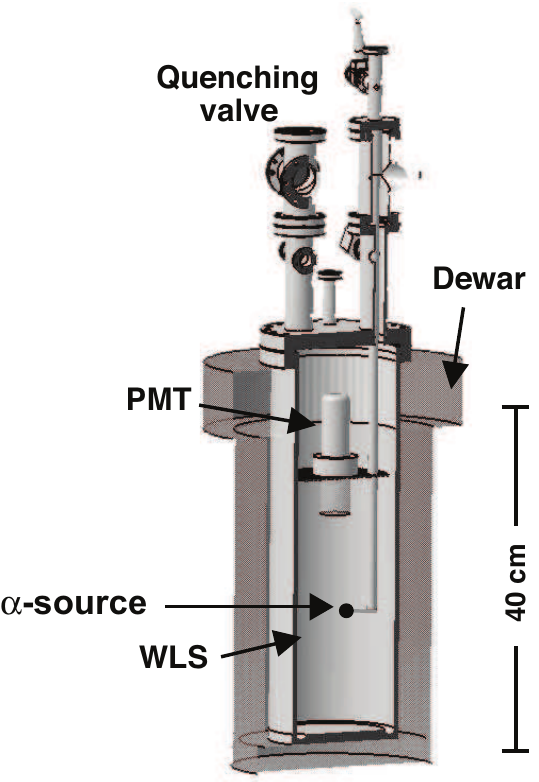}
    \caption{\small\it 6\,$\ell$ vessel for gaseous and liquid argon measurements.}
    \label{fig:LArsetup}
  \end{figure}

The apparatus consists of a 6\,$\ell$ cylindrical vessel,  150\,mm in inner diameter (with no internal electrical field), equipped with a variety of service connections, e.g. to a quadrupole mass spectrometer (fig.\,\ref{fig:LArsetup}).
Pumping is achieved with a dry primary pump and a 60\,$\ell$/s turbo pump. Without baking a typical vacuum of $10^{-5}$ mbar is reached within 24\,h. The residual gas is mainly composed of water vapour. The surrounding dewar is used for measurements with liquid argon. The present measurements are performed with a Hamamatsu R580 photomultiplier (PMT), 38 mm diameter with bialkali photocathode. A reflecting collar (MgF$_2$ coated aluminium)  concentrates the light on the photocathode. A 40\,Bq $^{210}$Pb source (half life  210 yrs), emitting 5.3\,MeV $\alpha$-particles   and up to 1.2\,MeV electrons ($^{210}$Pb $\to$ $^{210}$Bi $\to$ $^{210}$Po $\to$ $^{208}$Pb), is mounted in the center of the vessel, about 6 cm below the photomultiplier. The range of the $\alpha$-particles is roughly 4.5 cm in argon gas at NTP (density  1.78 g/l). Hence  $\alpha$-particles are fully absorbed while electrons from the source are not fully contained. The average number of 128 nm photons in pure argon gas is estimated to be 78'000, assuming an energy expenditure of about 68 eV/photon \cite{rico}.

\begin{figure}[htb]
 \centering
    \includegraphics[width=50mm]{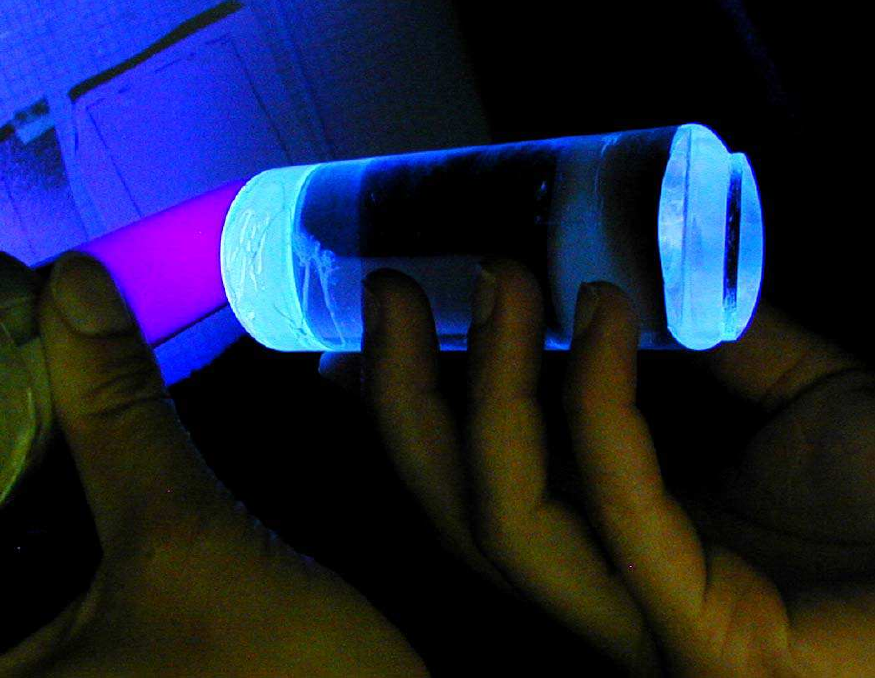}
    \caption{\small\it Trapping of the shifted blue light in a plexiglas cylinder.}
    \label{WLSspectrum}
  \end{figure}

Argon gas is taken from 50\,$\ell$ cylinders of class Ar\,60 (impurities\,$\leq$\,1.3\,ppm). An  oxisorb (CuO) filter and a hydrosorb  cartridge (molecular sieve) are used during filling.  The VUV scintillation light is converted into  blue light with  tetraphenyl-butadiene (TPB) as a wavelength shifter (WLS), sprayed on Tetratex or 3M (ESR vikuiti) foils. The emission spectrum of TPB reaches its maxi\-mum value at 422 nm \cite{Starnabrand} and matches the response of bialkali photocathodes (300 -- 600 nm). The typi\-cal time constant of TPB is 2 ns \cite{Huang}.  The  WLS material is dissolved in chloroform and sprayed (or evaporated) on reflecting foils covering the internal wall of the vessel. To improve on the overall light collection yield the glass window of the phototubes can also be coated with WLS. The coating is performed with TPB imbedded in a clear polymer matrix such as polystyrene or paraloid. Figure \ref{WLSspectrum} demonstrates  trapping of the shifted blue light in a plexi\-glas cylinder, one end of which has been covered with TPB/polystyrene. 

The R$580$ PMT is operated at  --1'350 V bias voltage yielding a gain of roughly $10^{6}$. Its signal is
amplified by a factor of 10 in a fast NIM amplifier and fed to the 8 bit FADC of a LeCroy WP7100 digital oscilloscope terminated with 50$\Omega$. The signal  is sampled at a rate of 1 GHz and, for each event,  20'000 samples covering 20 $\mu$s are recorded and stored in compressed MatLab file format to hard disk, in packages of about 1000 events.  

\begin{figure}[htb]
 \centering
    \includegraphics[width=65mm]{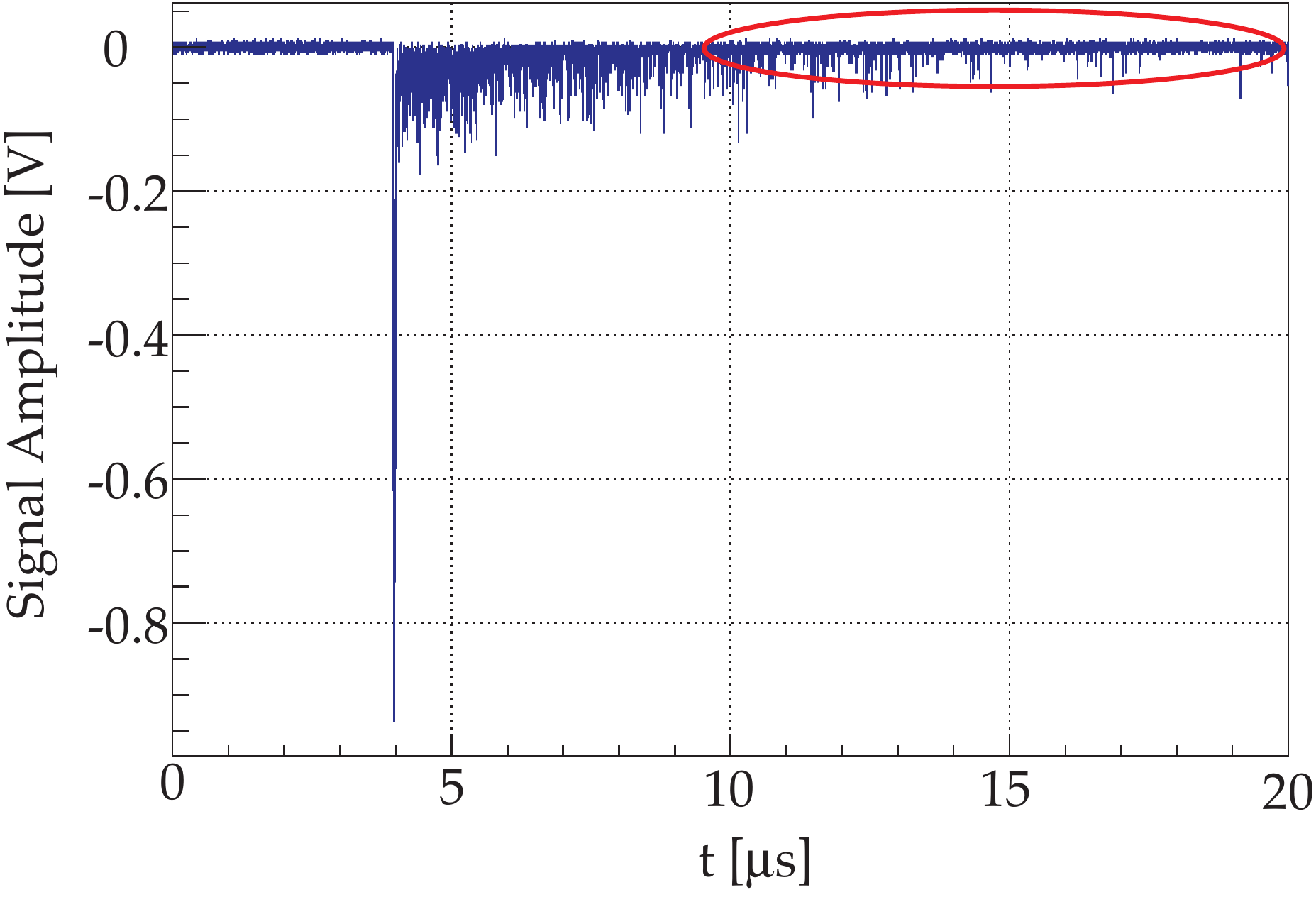}
    \caption{\small\it Sampling of an $\alpha$-event in gaseous argon. The ellipse shows the region dominated by single photon pulses.}
    \label{Evtsample}
\end{figure}

\section{Measurement of the decay spectrum}
\label{measure}

The measurements presented in this paper were collected at an argon pressure of 1100 mbar and at room temperature, except for the data at the partial air pressure of 2$\times$10$^{-6}$ mbar, which were taken slightly below the water freezing point  ($\sim$--20$^\circ$C). The latter constitute our cleanest argon sample. A typical event generated by the $\alpha$-source   at  2$\times$10$^{-6}$ mbar is shown in fig. \ref{Evtsample}. The large prompt pulse is mainly due to the fast scintillation component, the slowly decaying amplitude to the slow component, and the pulse at large sampling times (ellipse) to late arriving  single photons. The event trigger in the oscilloscope is set on the analogue signal height in the range of $-20$ to $-2'000$ mV, depending on the data set to be taken. The typical height of a single photoelectron pulse after the $10\times$ amplification amounts to roughly $-35$ mV.  

\begin{figure}[htb]
 \centering
    \includegraphics[width=65mm]{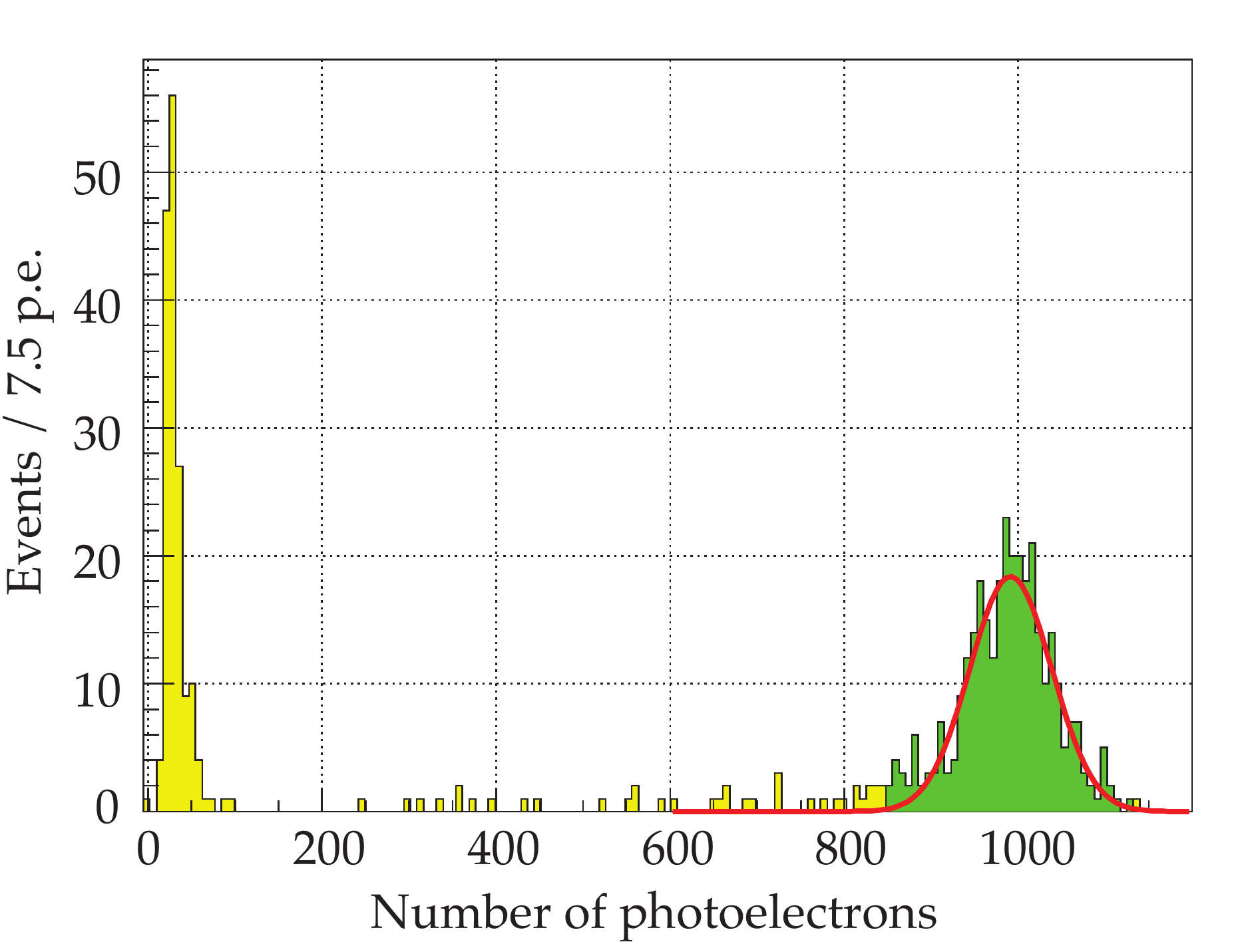}  
   \caption{\small\it Pulse height distribution of 321 $\alpha$-decays at a partial air pressure of 2 $\times$ 10$^{-6}$ mbar.}
    \label{Intpulse}
\end{figure}

Before integration the pedestal is determined event-wise by averaging over the first couple of thousand data points before the trigger. The total number of photoelectrons is then calculated by dividing the integrated signal (charge) by the average single photon charge, which we find from the integrated distribution  of dark counts. Single photon events from the signal tails  (ellipse in fig. \ref{Evtsample}) were used to for a cross-check only since they are affected by pile-up. Figure \ref{Intpulse} shows a typical  integrated pulse height  distribution for a sample of 528 events at a partial air pressure of 2 $\times$ 10$^{-6}$ mbar. The peak (321 events) is due to $\alpha$-particles, the counts below the peak to electrons. An integrated charge of 1 nVs over 50$\Omega$  corresponds to 6.8 $\pm$ 0.2   photoelectrons (p.e.).

\begin{figure}[htb]
 \centering
    \includegraphics[width=70mm]{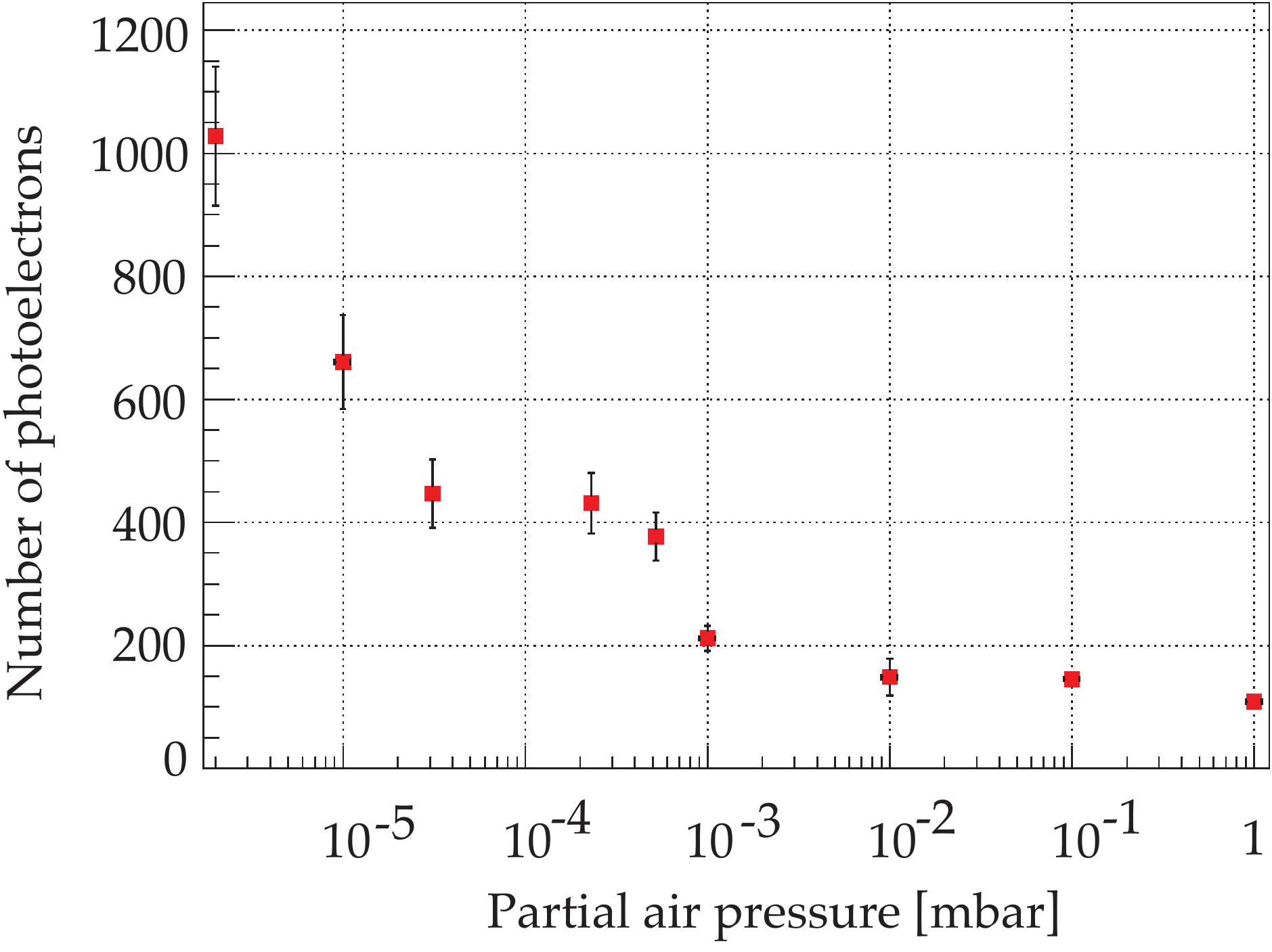}
\caption{\small\it Light yield of $\alpha$-particles in argon gas vs. partial air pressure at room temperature. The data at 2 $\times$ 10$^{-6}$ mbar were taken below the water freezing point  ($\sim$--20$^\circ$C).}
    \label{lyvsairnew}
\end{figure}

We now discuss the light yield as a function of gas purity which depends on the vacuum pressure in the vessel before letting argon in. Measurements were started by evacuating the vessel down to pressures around   10$^{-3}$ mbar. Air was then introduced up to atmospheric pressure and pumped out until the desired pressure was achieved. Gaseous argon was then introduced until a pressure of 1'100  mbar was reached.
Figure \ref{lyvsairnew} shows  the number of photoelectrons as a function of partial air pressure.
The number of photoelectrons clearly depends on the purity of the argon and still rises below 10$^{-5 }$ mbar.
The minimum pressure that could be achieved in the vessel was around $\sim$10$^{-6 }$ mbar by cooling with cold nitrogen.

\begin{figure}[htb]
 \centering
    \includegraphics[width=60mm]{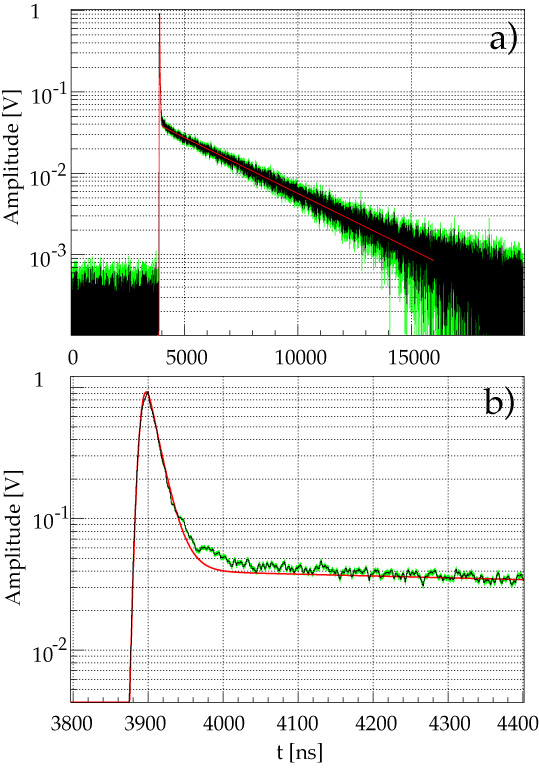}
\caption{\small\it a) Average signal from $\alpha$-decays in clean argon gas; b) zoom into the first 400 ns (see text).}
    \label{Cleangas}
\end{figure}

Next, we add the sampled pulse distributions from many $\alpha$-decays with total charge within $\pm 3\sigma$ from the peak value (see fig. \ref{Intpulse}), and determine the average charge and its error for each 1 ns bin.  Figure \ref{Cleangas}a shows a semilogarithmic representation of the  average amplitude vs. time for the 321 $\alpha$-events above, taken at  2 $\times$ 10$^{-6}$ mbar. The dark (black) area represents the data points, the grey (green) area the r.m.s errors.  Figure \ref{Cleangas}b shows a zoom on the first 400 ns 

The red line shows the fit to the data with two exponentials to model the singlet (fast) and triplet (slow) excimer decays, convoluted with a gaussian $G$ to describe the experimental resolution. A step function $H$ and a free parameter are used to fit the event time $t_{0}$. Hence the model function has 6 free parameters and is given by

\begin{eqnarray}
f & = &
G(t-t_0,\sigma) \nonumber\\
& \otimes &
H(t-t_{0}) 
\left[\frac{A}{\tau_{1}}  e^{-\frac{t-t_{0}}{\tau_{1}}}
+\frac{B}{\tau_{2}}   e^{-\frac{t-t_{0}}{\tau_{2}}}\right],
\label{eq1}
\end{eqnarray}
where $H(t-t_{0})$ = 1 for $t>t_0$ and 0 for $t<t_0$.   The parameters $\tau_1$ and $\tau_2$ are the time constants  of the singlet and triplet states. The parameters $A$ and $B$ are proportional to the total number of photoelectrons detected from the singlet, respectively  triplet states. In the absence of non-radiative de-excitation processes, they   des\-cribe  the populations of the two excimer states. 

Decays from both states are clearly visible in fig. \ref{Cleangas}. The fit is in good agreement with data, except in the intermediate region ($t-t_0\sim$150 ns). The slight excess of data in this region (possibly due to an additional production process) does not affect our results on the singlet and triplet time constants. We have fitted a third exponential ($\tau_3$ $\sim$100 ns) and also performed fits without the 3.94 -- 4.20 $\mu$s region. We use these fits to assess the (dominating) systematical errors. For the data shown in fig. \ref{Cleangas} we find $\tau_1$ = 14.7 $\pm$ 2.5 ns, $\tau_2$ = 3'140 $\pm$ 67 ns and a ratio $R \equiv B/A$ = 5.5 $\pm$ 0.6. The fitted r.m.s. resolution $\sigma$ of 5.6 ns is consistent with the time constant of the PMT (4.5 ns), including the transit time of $\alpha$-particles ($\sim$4ns).

Our result for $\tau_2$ is in good agreement  with a previous measurement  in dense gases, 3.2 $\pm$ 0.3 $\mu$s \cite{keto}, but is more precise. Our value for $\tau_1$ includes production and decay times of the singlet state. We have therefore performed a further fit, replacing  in eqn. (\ref{eq1}) the square bracket  by 
\begin{equation}
\left[\frac{A}{\tau^p_{1}-\tau^d_{1}} \left( e^{-\frac{t-t_{0}}{\tau^p_{1}}}-e^{-\frac{t-t_{0}}{\tau^d_{1}}}\right)+\frac{B}{\tau_{2}}   e^{-\frac{t-t_{0}}{\tau_{2}}}\right]
\end{equation}
where $\tau^p_{1}$ is the production time for  the singlet state, while $\tau^d_{1}$ is its mean life.  We find that $A$, $B$ and $\tau_2$ do not change significantly and  that $\tau^d_{1}\ll\tau^p_{1}$. Fixing $\tau^d_{1}$ to 4 ns \cite{keto} leads to a production time $\tau^p_{1} \simeq$ 14 ns   $\simeq \tau_1$. Hence  the time constant of the singlet state is determined  by its production mechanism. Our value for $\tau_1^p$ with $\alpha$-particles is smaller than  with electrons ($\simeq$40 ns at 3 atm \cite{keto}).

\begin{table}[htb]
\caption[]{\small\it Time  constant $\tau_1$ and  mean life $\tau_2$  for  various residual air pressures at room temperature. $A$ and $B$ are  proportional to the   total number of photoelectrons from each component.}
\vspace{5mm}
\begin{center}
\begin{tabular}{r|rr|rr}
\hline
p [mbar] & $\tau_1$ [ns]  & A [nVs]  
 & $\tau_2$ [ns] & B [nVs]  \\
 \hline
1 &  \  $\simeq$10    & 9.5 (0.2)  
&  \ 114 (9) & \ 6.5  (0.4) \\
10$^{-1}$ & \ $\simeq$10  &  12.7  (0.3)  
 & \ 198 (14) &   \ 8.7  (0.4) \\
10$^{-2}$ & \ 11.0 (0.6)    & \ 12.8   (0.3)  
 &\  240 (57) & \  9.1  (1.8) \\
10$^{-3}$ & \ 11.9 (1.7)  & \ 14.0  (0.8) 
  &  \ 520 (59) & \ 17.1  (1.2) \\
5.2 $\times$ 10$^{-4}$ & \ 12.7 (2.8)  & \ 15.6 (1.5) 
 & \ 1'341 (44) &  \ 39.9  (0.8) \\
2.3 $\times$ 10$^{-4}$ & \ 12.8 (2.8)  & \ 14.8 (1.6)  
& \ 1'788 (43) &  \ 48.6  (0.7)  \\
3 $\times$ 10$^{-5}$ & \ 13.2 (2.9)    & \ 15.2 (1.8)   
   & \ 1'817 (44) &  \ 50.5  (0.6)  \\
10$^{-5}$ & \ 12.6 (2.8)   & \ 14.5 (1.6)   
  & \ 2'992 (33) & \ 82.6  (0.4)  \\
\hline
\end{tabular}
\vspace{5mm}
\label{Tabpress}
\end{center}
\end{table}

\begin{figure}[htb]
 \centering
    \includegraphics[width=75mm]{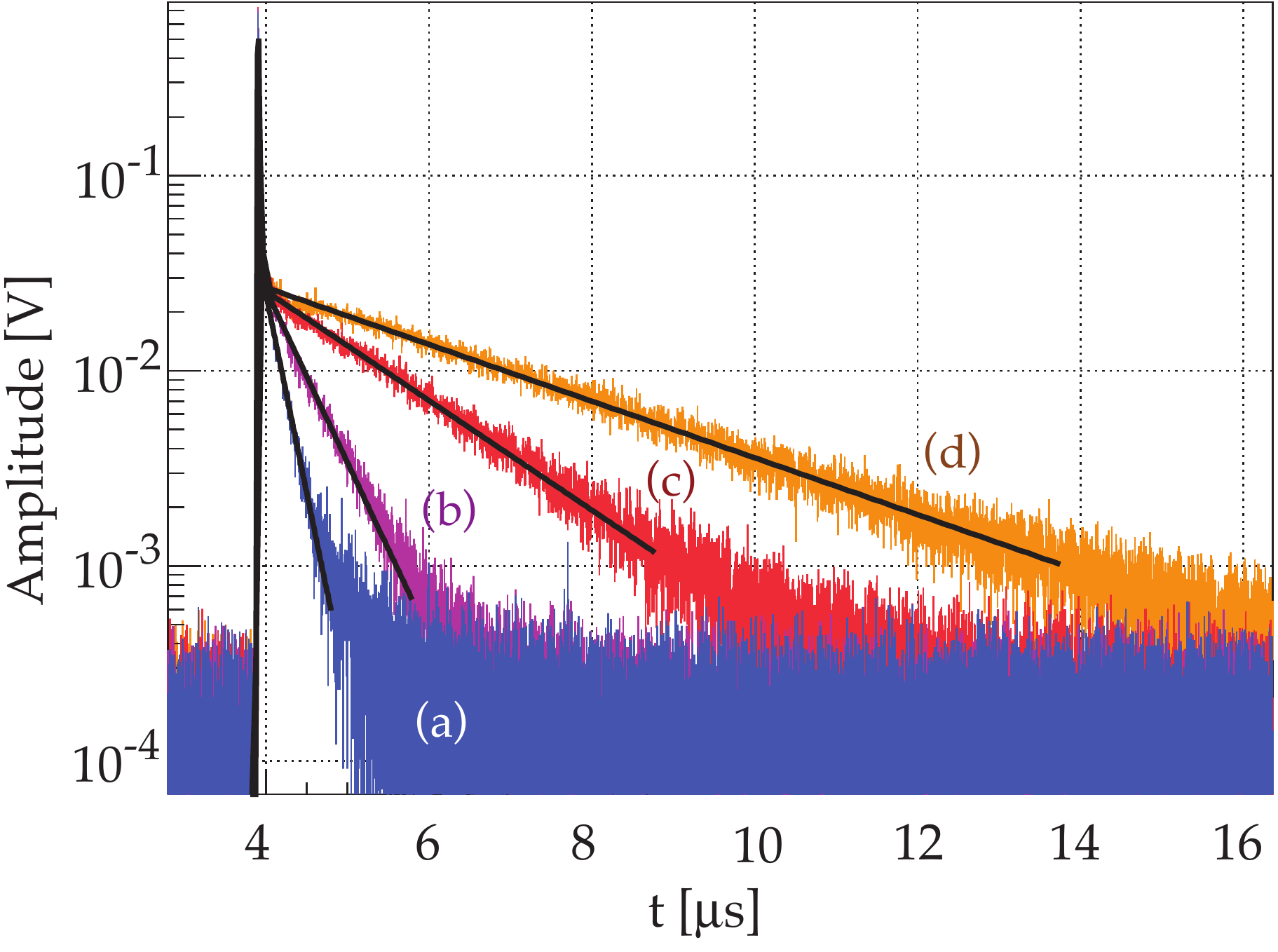}
    \caption{\small\it Average signal from $\alpha$-decays for various residual air pressures: 10$^{-2}$ mbar (a), 10$^{-3}$ mbar (b), 2.3 $\times$ 10$^{-4}$ mbar (c) and 10$^{-5}$ mbar (d).}
    \label{Ampvspress}
\end{figure}

The time distributions for vari\-ous  partial air pressures and the corresponding fits using eqn. (\ref{eq1}) are shown in  fig. \ref{Ampvspress}. The  time constants $\tau_1$, $\tau_2$ and the amplitudes $A$ and $B$  are given in Table \ref{Tabpress}.
The time constant of the fast component and its photoelectron yield remain constant within errors. From Table \ref{Tabpress} the average value for $\tau_1$ is 11.3 $\pm$ 2.8 ns. In contrast, the mean life $\tau_2$ of the slow component and its photoelectron yield clearly increase with gas purity. 

For each measurement, two batches of data were taken consecutively, the second
about 10 -- 15 min later than the first. The number of photoelectrons decreased slightly between the
first and the second batches. However, with cooling, the second batch contained slightly more photoelectrons. This  suggests in the first case outgassing of water molecules  and, in the second case, freezing water on the walls of the vessel. Hence the partial pressure of water might be responsible for the non-radiative destruction of the triplet states. 

\section{Triplet to singlet production ratio}
\begin{figure}[htb]
    \centering
    \includegraphics[width=75mm]{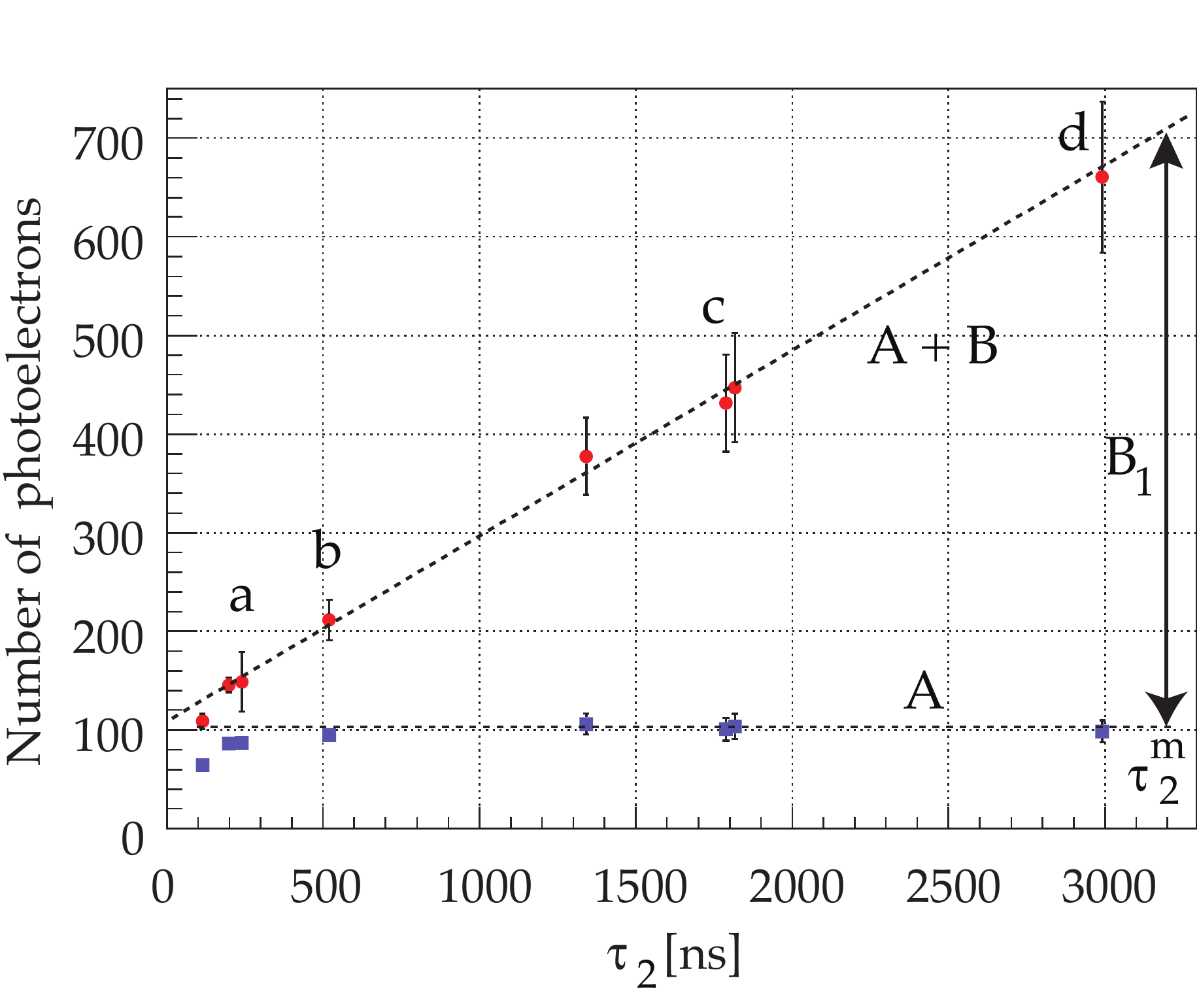}
    \caption{\small\it Total number of photoelectrons  as a function of $\tau_2$. The upper dashed line  gives the fitted total light yield ($A + B$) from both components, the bottom dashed line  the singlet contribution (A). The labels a -- d refer to the data shown in fig. \ref{Ampvspress}.}
    \label{fig:lyvs2ndt}
\end{figure}

Since $\tau_1$ and $A$ do not depend on gas purity, we can determine the ratio of triplet to singlet po\-pulations  by measuring the light yield as a function of $\tau_2$ and extrapolating to pure argon gas. Indeed, the light yields being proportional to $A$ and $B$, the population $B_1$ of the triplet state
 is given by $B = B_1 \Gamma_\gamma \tau_2$, 
where $\Gamma_\gamma$ is the radiative width, while the population of the singlet state is proportional to $A$.  The ratio of triplet to singlet populations is then equal to the ratio $R=B/A$ for $\Gamma_\gamma$ = $1/\tau_2$, i.e. for pure argon gas.  Figure \ref{fig:lyvs2ndt} shows the total number of photoelectrons as a function of $\tau_2$  for the various gas purities given in Table \ref{Tabpress}. Extrapolating $\tau_2$ to $\tau_2^m$ = 3.2 $\mu$s \cite{keto} we find $R = 5.5 \pm 0.6$  for 5.3 MeV $\alpha$-particles in clean argon gas at 1100 mbar and room temperature. 

\begin{figure}[htb]
 \centering
    \includegraphics[width=75mm]{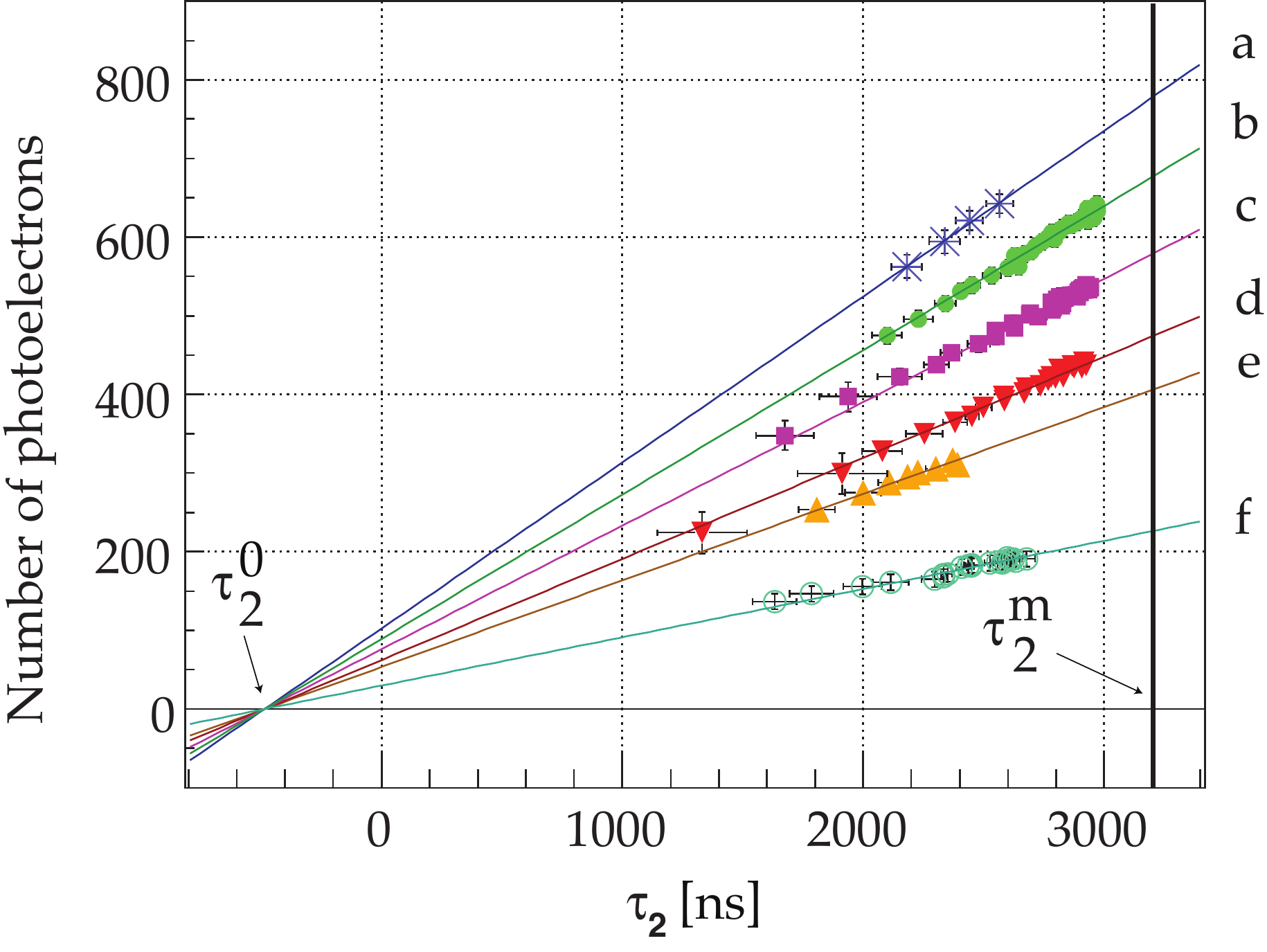}
    \caption{\small\it Light yield of $\alpha$-particles in argon gas for different  WLS coating and application techniques:  
    a) coated PMT and sprayed  3M-foil; b) sprayed PMT and 3M-foil; c) sprayed 3M-foil ; d) sprayed PMT ($\sim$0.4 mg/cm$^2$) and   Tetratex; e) sprayed PMT ($<$0.1 mg/cm$^2$) and  Tetratex; f) sprayed PMT  and MgF$_2$ coated mirror foil.}
    \label{fig:combplot}
\end{figure}

We have also performed measurements of the light yield for a series of WLS configurations. These data provide a cross-check for the measurements presented above. The PMT
photocathode was either coated with TBP/polystyrene or sprayed with TPB, and the reflecting foil (3M or Tetratex) sprayed with TPB of various thicknesses. Figure \ref{fig:combplot} shows the various measurements. As in fig. \ref{fig:lyvs2ndt} we extrapolate the data points to the expected decay time of $\tau_2^m$ = 3.2 $\mu$s for the second component in pure argon. However, we now require the straight lines to intersect  the $\tau_2$-axis at the common fitted point $\tau_2^0$ = --485$^{+100}_{-150}$ ns, where $A+B$ = 0. The simple proportionality $R = B_1/A$ = $\tau_2^m /\tau_2^0$ leads to the population ratio $R$ = 6.6$^{+2.1}_{-1.5}$.  This number agrees with our determination above, but is much less precise due to the large lever arms in the extrapolation to $\tau_2^0$.

Summarizing, we have measured the light yield in gaseous argon for various residual air pressures. For the slow component of the luminescence at 128 nm we observe a strong dependence on resi\-dual air pressure. This effect is attributed to contami\-nating water vapour. The longest  mean life is obtained from purest argon (3'140 $\pm$ 67 ns). The population ratio between the slow (spin-triplet) and the fast (spin-singlet) excimer states is measured for $\alpha$-particles to be $R$ = 5.5 $\pm$ 0.6 in pure argon gas at 1100 mbar and room temperature.\\ 

\section*{Acknowledgments}

We thank Hugo Cabrera and Andreas Knecht for their contributions in developing the reflector foils and wavelength shifters, Marco~Laffranchi for preparing the radioactive source, and the ArDM collaboration for valuable discussions and comments.
This work was supported by a
grant from the Swiss National Science Foundation.\\

\end{document}